\documentclass{jpp}
\usepackage[T1]{fontenc}
\usepackage[utf8]{inputenc}
\usepackage{amsmath}
\usepackage{graphicx}
\usepackage{babel}
\usepackage{epstopdf, epsfig}

\newcommand{\replace}[2]{#2}

\usepackage{subcaption}

 \usepackage{amssymb}
 \usepackage{bigints}
 \usepackage{empheq}
 \usepackage{framed}
 \usepackage{color}
 \usepackage{enumerate}
 \usepackage{ulem}
 \usepackage{bm}
 \usepackage{appendix}
 \numberwithin{equation}{section}
 \usepackage{upgreek}
 \usepackage{textgreek}
 \usepackage{graphicx}
\usepackage{stmaryrd} %llbrackets
\usepackage{xfrac}    % Works better with other fonts

\newcommand{\be}{\begin{equation}}
\newcommand{\ee}{\end{equation}}
\newcommand{\ba}{\[\begin{aligned}}
\newcommand{\ea}{\end{aligned}\]}
\newcommand{\bea}{\begin{eqnarray}}
\newcommand{\eea}{\end{eqnarray}}
\newcommand{\beann}{\begin{eqnarray*}}
\newcommand{\eeann}{\end{eqnarray*}}
\newcommand{\bs}{\begin{split}}
\newcommand{\es}{\end{split}}

\newcommand*{\mff}{\mathfrak{f}}
\newcommand*{\mfg}{\mathfrak{g}}

\newcommand*{\B}{\bm{B}}

\newcommand*{\del}{\partial}

\newcommand*{\dlr}{\bm{\nabla_\perp}}
\newcommand*{\dlrs}{\nabla_\perp^2}

\newcommand*{\lbr}{\left(}
\newcommand*{\rbr}{\right)}

\newcommand*{\Pb}{\overline{P}}
\newcommand*{\Qb}{\overline{Q}}

\newcommand*{\Gammab}{\overline{\Gamma}}
\newcommand*{\gammab}{\overline{\gamma}}

\newcommand*\at[2]{\left.#1\right|_{#2}}

\newcommand*\reallywidetilde[1]{\ThisStyle{%
  \setbox0=\hbox{$\SavedStyle#1$}%
  \stackengine{-.1\LMpt}{$\SavedStyle#1$}{%
    \stretchto{\scaleto{\SavedStyle\mkern.2mu\sim}{.5467\wd0}}{.7\ht0}%
  }{O}{c}{F}{T}{S}%
}}

\def\test#1{$%
  \reallywidetilde{#1}\,
  \scriptstyle\reallywidetilde{#1}\,
  \scriptscriptstyle\reallywidetilde{#1}
$\par}
\makeatletter
\newsavebox{\@brx}
\newcommand{\llangle}[1][]{\savebox{\@brx}{\(\m@th{#1\langle}\)}%
  \mathopen{\copy\@brx\mkern2mu\kern-0.9\wd\@brx\usebox{\@brx}}}
\newcommand{\rrangle}[1][]{\savebox{\@brx}{\(\m@th{#1\rangle}\)}%
  \mathclose{\copy\@brx\mkern2mu\kern-0.9\wd\@brx\usebox{\@brx}}}
\makeatother

\usepackage{blindtext}

\title{A class of high-beta, large-aspect-ratio quasiaxisymmetric Palumbo-like configurations}

\author{Andrew Brown\aff{1}\aff{2}\corresp{\email{aobrown@princeton.edu}},
Wrick Sengupta\aff{1},
Nikita Nikulsin\aff{3},
\and Amitava Bhattacharjee\aff{1}}
\affiliation{\aff{1}Department of Astrophysical Sciences, Princeton University, Princeton, NJ 08544, USA
\aff{2}Plasma Physics Laboratory, Princeton University, Princeton, NJ 08543, U.S.A
\aff{3}Max Planck Institute for Plasma Physics, 17491 Greifswald, Germany}

\begin{document}
\maketitle
\begin{abstract}
The space of high-beta, approximately quasiaxisymmetric, large-aspect-ratio stellarator configurations is explored using an inverse coordinate approach and a quadratic polynomial ansatz for the flux function, following the method of Palumbo, extended by Hernandes and Clemente. This approach yields a system of nonlinear ODEs that, when solved, give equilibria exhibiting positive or negative triangularity, cusps, and (in an extreme limit) current singularities. It is shown that a cubic ansatz may also be used, but that polynomials of degree four or higher will lead to overdetermination.
\end{abstract}

\section{Introduction}
Omnigeneity \replace{}{\citep{nuhren1988, hall1975, cary1997}}, zero-average radial drift of the guiding center orbits, is essential for particle confinement in three-dimensional (3D) magnetic fields \citep{helander2014}. One route to omnigeneity is quasisymmetry \replace{(QS)}{(QS, \citep{boozer1983})}, in which contours of the magnetic field strength are flat in a special coordinate system known as \textit{Boozer angles}. QS with \replace{magnetohydrodynamic (MHD) force balance}{magnetohydrostatic (MHS) force balance, assuming isotropic pressure,} is known to be an overdetermined system \citep{garrenboozer1991a}. However, recent strides have shown that extremely precise QS can be obtained by numerical optimization \citep{landreman2021, wechsung_PNAS2022preciseQA_coils}, reinforcing the confidence that QS is a practical approach to plasma confinement. Because quasisymmetric equilibria have been found only numerically, our understanding of the qualitative behavior of quasisymmetric equilibria is still inchoate. Considerable effort has been applied to the near-axis theory of QS \citep{garrenboozer1991a, garrenboozer1991b, landreman2019ii, rodriguez2021b, rodriguez2021c} and to weakly nonaxisymmetric equilibria at finite aspect ratio \citep{plunk2018,henneberg2019,plunk2020_near_axisymmetry_MHD}. Relatively little work has been done to understand QS analytically far from the axis \replace{}{ \citep{sato2022}}. One reason for this gap is the lack of a standard Grad-Shafranov equation (of the elliptic type)\replace{}{, }well known for axisymmetric equilibria. In 3D, even assuming the existence of nested surfaces, the Grad-Shafranov equation is non-standard and has mixed characteristics \replace{}{\citep{burby2020}}.

In this work, we consider the limit of large aspect ratio and weak perturbations away from axisymmetry to extract a tractable, semi-analytical model that reproduces many of the qualitative features of numerically optimized QA configurations. The regime explored here is inaccessible by near-axis theory and difficult to analyze using \replace{VMEC}{the Variational Moments Equilibrium Code (VMEC, \cite{hirshman1983})} or similar tools because of sharp features in the plasma boundary \replace{}{which appear in many models, but in general are difficult to resolve with a Fourier representation \citep{schuett2024, gaur2025}}. We therefore stress the complementary role that analytical theory must play in enhancing, explaining, and providing new directions for purely numerical optimization.

Our approach is stimulated by a recent 3D quasisymmetric, high-beta stellarator (QS-HBS) model developed in \cite{Sengupta_Nikulsin_Gaur_Bhattacharjee_2024} and extended in \cite{Nikulsin_Sengupta_Jorge_Bhattacharjee_2024}, \replace{which obtains a Grad-Shafranov equation (of the elliptic type) by imposing asymptotic ordering}{based on the HBS model of Freidberg \citep{freidberg2014idealMHD}. Freidberg's model assumes large aspect ratio (so inverse aspect ratio $\epsilon \ll 1$) and ``high" plasma beta $\beta \equiv 2 \mu_0 B^2/p \sim \epsilon$, (with scalar thermal pressure $p$ and magnetic field $B$). The magnetic field, normalized by its value $B_0$ at the magnetic axis, is toroidal at $O(1)$, with an $O(\epsilon)$ poloidal part generated by the toroidal current. The QS-HBS model of \cite{Sengupta_Nikulsin_Gaur_Bhattacharjee_2024} imposed QA on Freidberg's HBS model and further required the magnetic axis to be a circle of radius $R_0 \equiv 1/\epsilon$ plus an $O(1)$ deviation from axisymmetry (where all distances are in units of the minor radius), and showed that the QA error is $O(\epsilon^2)$ if the non-axisymmetric perturbation to the axis is purely vertical. The generalization in \cite{Nikulsin_Sengupta_Jorge_Bhattacharjee_2024} showed that the axis may also have some radial perturbation so long as the axis torsion is $O(\epsilon^2)$, and the QA error then remains $O(\epsilon^2)$ with zero flux surface shaping. With all these conditions, the QS-HBS model relaxes the QS overdetermination problem and yields a Grad-Shafranov equation of elliptic type.}

In section \ref{sec:Extended_Palumbo}, we use this model to explore an interesting class of global solutions using a modified form of a method pioneered by Palumbo \citep{palumbo1968} and extended in \cite{bishop_Taylor1986degenerate,hernandes_Clemente2009ext_Palumbo} to construct isodynamic axisymmetric solutions \replace{ for which magnetic surfaces and constant-$|\B|$ contours coincide}{, i.e., flux surfaces on which $|\B|$ is constant}. In section \ref{sec:ext_palumbo_profiles}, we show that the resulting quasi-axisymmetric configurations exhibit many of the qualitative features observed in finite-aspect-ratio systems such as those found numerically by Landreman and Paul \citep{henneberg2019,landreman2021}. These include very low magnetic shear, D-shaped equilibria for small $\iota$, and inboard cusp formation as the maximum plasma volume is approached. \replace{}{We stress that these features of our model cannot be understood solely from the existing near-axis framework. Cusp formation in our model appears as the plasma reaches the maximum allowable volume, which differs from surface self-intersection in the near-axis theory, and which simply indicates the breakdown of the near-axis approximation \citep{landreman2021a}. Similarly, the magnetic shear can be calculated from the near-axis theory \citep{rodriguez2023}, but such calculations are only strictly valid in the vicinity of the axis.} In section \ref{sec:axis_pert}, we exploit the decoupled axis and plasma shaping to show that non-negligible rotational transform may be generated from weak axis perturbations, and we use VMEC to verify the asymptotic behavior of the QA error. We conclude by considering some possible applications of this work to quasi-axisymmetric stellarator design.

\section{Analytic solutions of the Quasisymmetric Grad-Shafranov Equation: An extended Palumbo approach}\label{sec:Extended_Palumbo}
Throughout this work, we use standard cylindrical coordinates $(R, \phi, Z)$ and local Cartesian coordinates $(x, \xi)$ centered on the magnetic axis $(R_0(\phi), Z_0(\phi))$
such that $x=R - R_0(\phi)$ and $\xi = Z- Z_0(\phi)$\replace{}{, and define the perpendicular gradient $\nabla_{\perp}\equiv \left(\partial_x \text{, }\partial_{\xi}\right)^T$.}The quasisymmetric Grad-Shafranov equation (QS-GSE) of the QS-HBS model takes the form (in normalized units)

\begin{equation}
    \nabla_{\perp}^2 \Psi + 2x \frac{d \beta}{d \Psi} + \frac{d H}{d \Psi} = 0\replace{}{.}
    \label{eq:QS-GSE}
\end{equation}
and the toroidal current  \replace{}{ density $J_{\phi}$}, in terms of the profiles $\beta(\Psi)$ and $H(\Psi)$, is

\begin{equation}
    \frac{\replace{\mathbf{J_{\phi
    }}}{J_{\phi}}}{B_0} = -\epsilon \left(2x \frac{d \beta}{d \Psi} + \frac{d H}{d \Psi}\right).
    \label{eq:curr}
\end{equation}

We shall now employ an alternative approach based on an extension \citep{hernandes_Clemente2009ext_Palumbo, bishop_Taylor1986degenerate} of Palumbo's work on isodynamic equilibrium \citep{palumbo1968} to obtain exact solutions to the QS-GSE equation \eqref{eq:QS-GSE}. Our approach uses $(x,\Psi)$ as independent variables and solves for $\xi(x,\Psi)$. Therefore, it is an inverse coordinate approach to solve the QS-GSE. The primary advantage of the inverse-coordinate approach stems from the fact that $\xi$ enters the QS-GSE only through the $\dlrs$ term since the pressure term and the current terms are only functions of $(x,\Psi)$. Moreover, this approach shows that the MHD profile functions cannot be arbitrary if we impose QS.

\subsection{An inverse-coordinate approach to solve the QS-GSE}\label{sec:Extended_Palumbo_inv_coord}
Using an ingenious method, Palumbo \citep{palumbo1968} obtained MHD equilibria with isodynamic magnetic fields, where the magnetic field strength is a constant on a flux surface, i.e., $|\B|=B(\psi)$. Extensions of Palumbo's approach were carried out by \cite{bishop_Taylor1986degenerate}, \cite{hernandes_Clemente2009ext_Palumbo} and \cite{hernandes2013_D_shaped}. 
 
Following the approach outlined in \citet{hernandes_Clemente2009ext_Palumbo}, we now carry out the inverse-coordinate analysis starting with the variable transform from $(x,\xi)$ to $(x,\Psi)$ using the identities
\begin{align}
  \replace{\at{\del_x}{y}}{\at{\del_x}{\xi}}= \at{\del_x}{\Psi} + u\; \replace{\del_\Psi}{\at{\del_\Psi}{x}} , \quad \at{\del_\xi}{x}= v\; \del_\Psi, \quad u\equiv \Psi_{,x}, \quad v\equiv \Psi_{,\xi}.
\end{align}
Since $(u,v)$ are components of $\dlr \Psi$, \replace{they must satisfy the consistency condition}{commutation of partial derivatives} $\del_\xi u = \del_x v$, which can be written in the following equivalent forms \citep{hernandes_Clemente2009ext_Palumbo}:
\begin{align}
\lbr \del_x + u\;\del_\Psi \rbr v^2 =2 v^2 \del_\Psi u \quad \Leftrightarrow \quad \del_x \lbr \frac{1}{v}\rbr + \del_\Psi \lbr \frac{u}{v}\rbr=0.
    \label{eq:consistency_uv}
\end{align}
\replace{The QS-GSE \eqref{eq:QS-GSE} in terms of $u,v$ reads}{We proceed to solve the QS-GSE \eqref{eq:QS-GSE}, treating $u$ and $v$ as the unknowns. It now reads}
\begin{align}
\del_x u + \del_\Psi \Lambda =0
    \label{eq:GS_uv}
\end{align}
where, we have defined 
\begin{align}
\Lambda \equiv \frac{1}{2}\eta+2 x\beta(\Psi) +H(\Psi), \quad \eta(x,\Psi)\equiv u^2+v^2.
    \label{eq:Lambda_eta_def}
\end{align}
Once the QS-GSE equation \eqref{eq:GS_uv} is solved subject to the consistency condition \eqref{eq:consistency_uv}, we \replace{}{can} obtain $\xi$ \replace{through}{(and thus the flux surface geometry) by numerically integrating}
\begin{align}
    \xi = - \int \frac{u}{v}\; dx , \quad \text{or equivalently }\quad \xi =  \int \frac{1}{v}\; d\Psi.
\label{eq:xi_eq_uv}
\end{align}
The two integrals are equal to each other, as can be seen from the second form of the consistency condition \eqref{eq:consistency_uv}.

The system \eqref{eq:GS_uv}-\eqref{eq:Lambda_eta_def} is a highly nonlinear system of PDEs for $(u,v)$. Fortunately, analytical progress can be made based on the following observation. Integrating  \eqref{eq:GS_uv} in $x$, we get
\begin{align}
    u(x,\Psi)=u_0(\Psi)-x H'(\Psi)-x^2 \beta'(\Psi) -\frac{1}{2}\int dx\; \del_\Psi\eta(x,\Psi)
    \label{eq:u_form}.
\end{align}
The form of $u$ as given in \eqref{eq:u_form} suggests that one could seek a solution where $|\dlr\Psi|^2=\eta(x,\Psi)$ is a polynomial in $x$ with coefficients that are functions of $\Psi$, as done in \cite{hernandes_Clemente2009ext_Palumbo}. For the cases that \citet{palumbo1968} and \citet{bishop_Taylor1986degenerate} considered, the polynomial nature of $\eta$ is a direct consequence of the property of $|\B|$\replace{.}{, but in the model considered here it is an assumption made for tractability.}

We shall now seek a solution where $|\dlr\Psi|^2=\eta$ is a quadratic polynomial in $x$,
\begin{align}
    \eta(x,\Psi) = \eta_0(\Psi)+x\; \eta_1(\Psi) +x^2 \eta_2\replace{}{(\Psi)}.
    \label{eq:eta_ansatz}
\end{align}
Here $\eta_i, i=0,1,2$ are functions of $\Psi$ to be determined by solving the nonlinear system \eqref{eq:GS_uv}-\eqref{eq:consistency_uv}. The quadratic nature of $\eta$ is suggested by the fact that in \eqref{eq:u_form} the non-trivial terms that involve $\beta', H'$ are, at most, quadratic in $x$. Substituting \eqref{eq:eta_ansatz} into \eqref{eq:u_form} we get
    \begin{align}
    u(x,\Psi)&= u_0(\Psi) + x\: u_1(\Psi) +x^2 u_2(\Psi),\label{eq:u_form2}\
\end{align}
where,
\begin{align}
    u_1 &= -\lbr \frac{\eta'_0}{2}+H'\rbr, \quad 
    u_2 = -\lbr \frac{\eta'_1}{4}+\beta'\rbr\label{eq:u1_u2_def}. %\label{eq:u1_u2_u_forms}
\end{align}

We note that to maintain the quadratic nature of $u$, we have chosen $\eta_2$ to be a constant. 

With $\eta$ and $u$ available, we can now calculate $\Lambda, v^2$. From the definition of $\Lambda$ \eqref{eq:Lambda_eta_def}, we find that $\Lambda$ is a quadratic in $x$ of the form
\begin{align}
    \Lambda= \lbr\frac{\eta_0}{2} +H(\Psi)\rbr +2 x\lbr \frac{\eta_1}{4}+\beta(\Psi)  \rbr +\replace{\eta_2}{\frac{\eta_2}{2}}\: x^2
\end{align}
Next, from the forms of $\eta, u$ as given by \eqref{eq:eta_ansatz}, \eqref{eq:u1_u2_def}, we have
\begin{align}
v^2=\eta -u^2,
    \label{eq:vsqr_eta_reln}
\end{align}
which can be expressed in terms of the following quartic in $x$
\begin{subequations}
    \begin{align}
    v^2&= \sum_{n=0}^{n=4}V_n(\Psi) x^n,\\
    V_0&=\eta_0-u_0^2, \quad V_1=\eta_1-2u_0 u_1,\quad V_2= \replace{\eta_2}{\frac{\eta_2}{2}}-2 u_0 u_2 -u_1^2\\
    V_3 &=-2 u_1 u_2 ,\quad V_4 =-u_2^2.
\end{align}
\end{subequations}
We are now in a position to evaluate the consistency condition \eqref{eq:consistency_uv} that would help us determine the functions of $\Psi$ such as $u_0,u_1,\eta_0,\eta_1$ etc. Using $v^2=\eta-u^2$, the consistency condition can be written in the following alternative form 
\begin{align}
    \lbr \del_x + u \del_\Psi -2u_{,\Psi}\rbr \eta = \del_x u^2,
    \label{eq:consistency_eta_u}
\end{align}
which leads to a fourth-order polynomial in $x$. Equating each of the coefficients of the polynomials to zero, we get five nonlinear coupled ODEs.

Equating the coefficient of $x^4$ to zero we get the condition 
\begin{align}
    u_2=\text{constant}.
\end{align}
The nonlinear ODEs can be simplified by using the following variables inspired by \cite{bishop_Taylor1986degenerate}:
\begin{align}
    P=\frac{\eta_0}{u_2^2}, \quad Q=\frac{\eta_1}{u_2^2}, \quad X=\frac{u_1}{u_2}, \quad \Gamma=\frac{u_0}{u_2}, \quad \gamma=\replace{\frac{\eta_2}{u_2^2}}{\frac{\eta_2}{2 u_2^2}} ,\quad \beta_p=\frac{\beta}{u_2^2},\quad H_p=\frac{H}{u_2^2}.
    \label{eq:PQXetc_def}
\end{align}
Furthermore, \replace{we shall define a variable $\lambda$ such that}{the equations take a simpler form if we rescale the flux $\Psi$, so we define a new independent variable $\lambda \equiv \Psi/u_2$, so}
\begin{align}
    \frac{d}{d\lambda}= u_2 \frac{d}{d\Psi} \equiv \dot{( \;\; )}.
    \label{eq:lambda_def}
\end{align}
In these variables, equating to zero the coefficients of $(x^0,x,x^2,x^3)$ of the consistency condition \eqref{eq:consistency_eta_u}, leads to the following coupled nonlinear ordinary differential equations (ODEs) for the system $(P,Q,X,\Gamma)$
\begin{subequations}
    \begin{align}
    Q-2 X \Gamma + \Gamma \dot{P} -2 P \dot{\Gamma}&=0\\
    2\gamma -2 X^2 + \Gamma(\dot{Q}-4) -2 Q\dot{\Gamma} + X \dot{P}-2 P \dot{X} &=0\\
    X(\dot{Q}-6)+ \dot{P}-2(Q \dot{X} + \gamma\dot{\Gamma}) &=0\\
   \dot{Q}-4-2\gamma \dot{X}&=0
\end{align}
\label{eq:BTS_coupled_ODEs}
\end{subequations}
\replace{}{Admissible initial conditions for the system \eqref{eq:BTS_coupled_ODEs} are determined by enforcing regularity of $\Psi$ at the magnetic axis and requiring that $v^2$ have a double root at the axis. The details of the near-axis asymptotics may be found in Appendix \ref{app:Palumbo}. The crucial result is that all of the functions $(P, Q, X, \Gamma)$ are completely fixed by choice of two parameters: $X_0$ and $\gamma$.}
Once the system of ODEs \eqref{eq:BTS_coupled_ODEs} are solved \replace{}{ numerically, using the MATLAB routine ode23tb}, we can find $H,\beta$ from \eqref{eq:u1_u2_def} by integrating
\begin{align}
   \dot{H_p}= -\lbr \frac{1}{2}\dot{P} +X \rbr, \quad \dot{\beta_p} = - \lbr 1+ \frac{1}{4}\dot Q \rbr.
    \label{eq:beta_prime_exp}
\end{align}

Though $u_2$ appears to be a free parameter, if the leading-order magnetic field $B_0$ is normalized to unity, the poloidal field must be $O(\epsilon)$, which gives some restriction on $u_2$ \textit{viz.}
\begin{align}
    |\nabla \Psi| &= \sqrt{\eta} = \sqrt{\eta_2 x^2 + \eta_2 x+ \eta_0} \\
    &= |u_2| \sqrt{\replace{\gamma}{2 \gamma} x^2 + Q x + P},
\end{align}
so if the radicand is $O(1)$, we are forced to choose $u_2$ to be at most $O(\epsilon)$.

Next, we discuss how we can obtain $\xi$ using \eqref{eq:xi_eq_uv}. The denominator of the integrand is $v$, which is a square root of a fourth order polynomial in $x$. Therefore, the integration with $x$ can be carried out exactly in terms of elliptic integrals. It follows that
\begin{align}
    \xi = \pm X\int_{s_1(\Psi)}^s ds \frac{\Gamma/X^2+s(s-1)}{\sqrt{\lbr P/X^4-s Q/X^3+s^2 \gamma/X^2 \rbr-(s(s-1)+\Gamma/X^2)^2 }}, \quad s\equiv -\frac{x}{X},
    \label{eq:xi_exp}
\end{align}
where $s_1(\Psi)$ is a value of $s$ such that the denominator vanishes. 

It is possible to apply the same method assuming a polynomial ansatz of greater degree for $u$. It is shown in Appendix \ref{app:overdet} that this procedure may be carried out without issue for a cubic polynomial, but for polynomials of degree $4$ or greater, the resulting system of ODEs is overdetermined.

The choice of a quadratic ansatz for $u$ is \textit{post hoc} justified by the observation that, to lowest order in $\epsilon$,
\begin{equation}
    \mathbf{b}\cdot \nabla B = -\epsilon^2 \frac{\partial_{\xi} \Psi B_0}{(1+\epsilon x)^2} + O(\epsilon^3)
\end{equation}
and the field strength satisfies
\begin{equation}
    B = B_0(1-\epsilon x) + O(\epsilon^2),
\end{equation}
so if $v^2$ is quartic in $x$, $\left(\frac{d B}{d\ell}\right)^2$ will be a quartic in $B$ (where $\ell$ is the length along the field line). In \cite{sengupta2024KdV}, \replace{convincing evidence has been provided that}{thousands of QS configurations from the QUASR database \citep{giuliani2024} were found to share the property that}, in a quasiaxisymmetric field, $\left(\frac{dB}{d\ell}\right)^2$ must be either a cubic or a quartic polynomial in $B$. The weakly perturbed tokamaks of the QS-HBS model \citep{Sengupta_Nikulsin_Gaur_Bhattacharjee_2024, Nikulsin_Sengupta_Jorge_Bhattacharjee_2024} certainly allow deviation from this form for $\frac{dB}{d\ell}$, but the Palumbo-like configurations of \citep{bishop_Taylor1986degenerate, hernandes_Clemente2009ext_Palumbo} do take the
form 
\begin{equation}
    \left(\mathbf{B}\cdot \nabla B^2\right)^2 = \mathbb{P}_{3,4}(B^2)\replace{.}{,}
\end{equation}
\replace{}{where $\mathbb{P}_k(t)$ indicates an unspecified polynomial of degree $k$ in the variable $t$.}

Assuming a quartic form for $\frac{dB}{d\ell}$, we must have 
\begin{equation}
    v^2 = \eta - u^2 = \mathbb{P}_4(x),
\end{equation}
making the simple choice of quadratic polynomials for $\eta$ and $u$ especially appealing. 

This completes the inverse-coordinate approach to \replace{obtaining an}{the} analytic solution to the QS-GSE. The ODE system \eqref{eq:BTS_coupled_ODEs} needs to be solved to determine the profiles $H$ and $\beta$ that appear in the QS-GSE using \eqref{eq:beta_prime_exp}. Thus, we can clearly see that the two MHD profile functions, and hence the rotational transform profile, cannot be arbitrary if we impose QS and \replace{MHD}{MHS} force-balance just like in the isodynamic case \citep{bishop_Taylor1986degenerate,schief2003nested,hernandes_Clemente2009ext_Palumbo}. Finally, we need to address the issue of the closure of the flux surfaces. Nestedness \replace{}{of flux surfaces} is guaranteed by the consistency condition \eqref{eq:consistency_uv} but to ensure that flux surfaces are closed we need to impose additional conditions, which we discuss in Appendix \ref{app:Palumbo}.

\section{Extended Palumbo Profiles}
\label{sec:ext_palumbo_profiles}

\begin{figure}
    \centering
    \includegraphics[width=0.6\columnwidth]{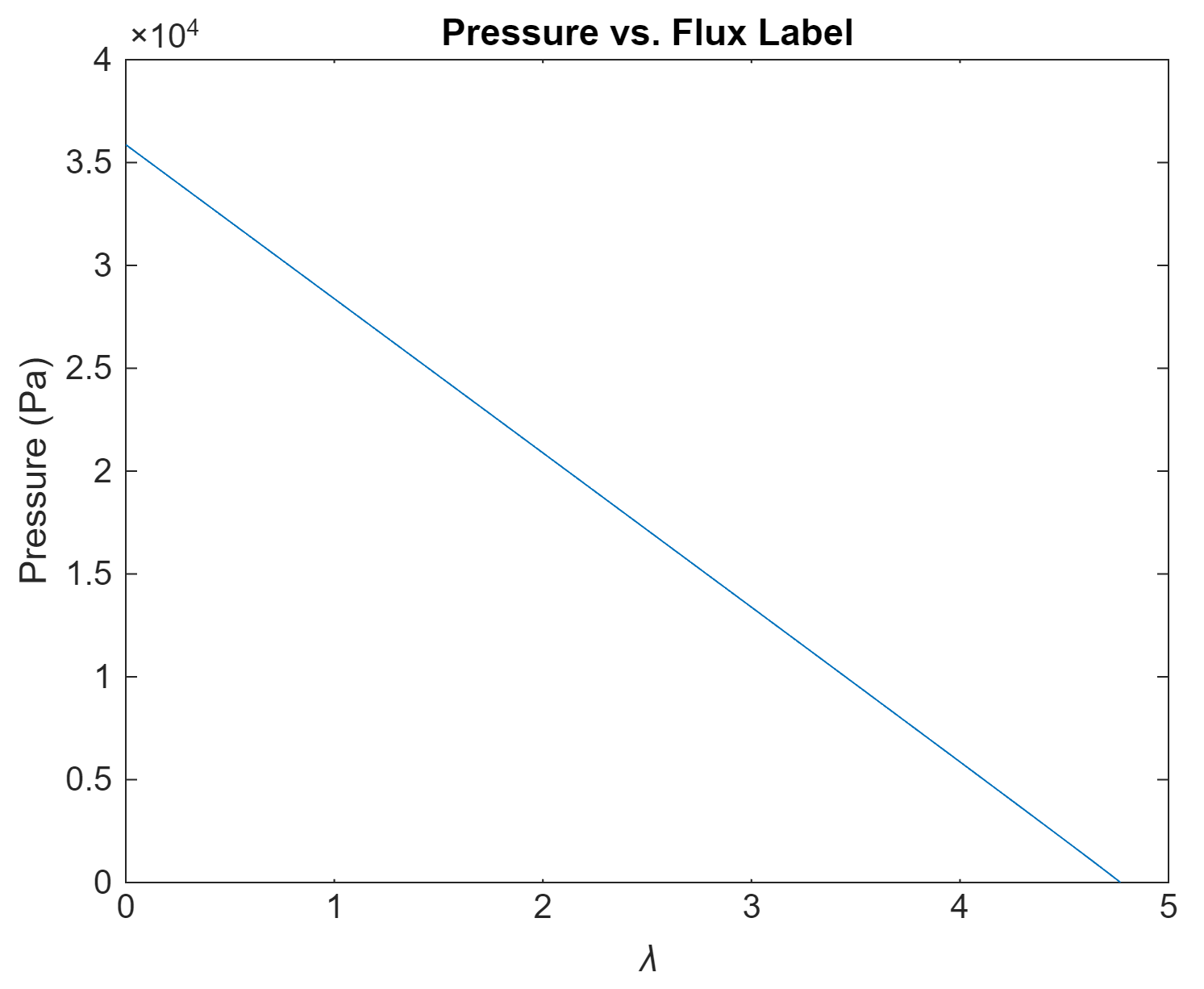}
    \caption{The pressure profile is nearly linear in the flux label, independent of the choice of parameters. Here $X_0 = 2.5$, $\gamma = 0.2$, and $\epsilon = 1/6$.}
    \label{pressure}
\end{figure}

\begin{figure}
    \centering
    \includegraphics[width=0.6\columnwidth]{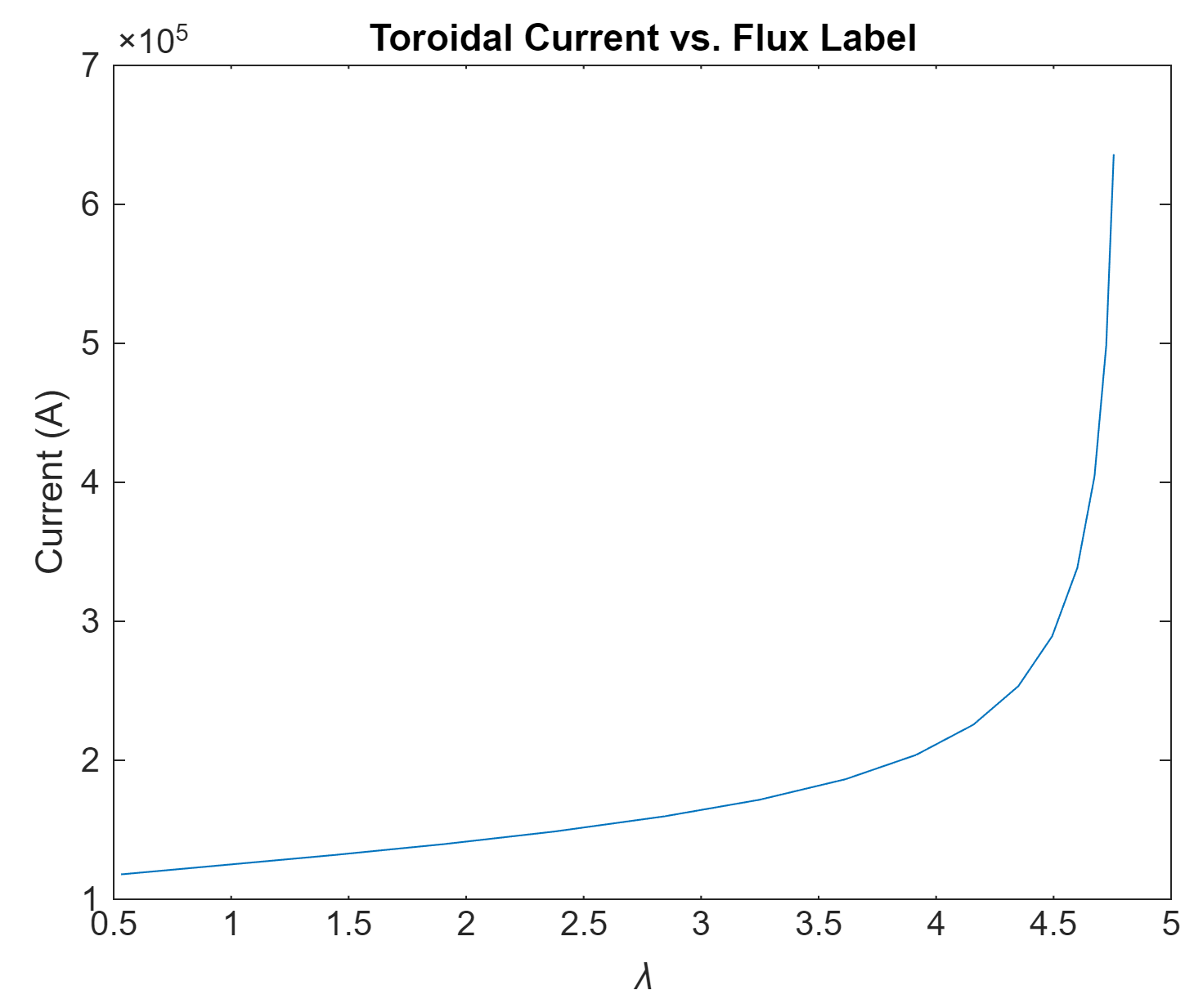}
    \caption{A current singularity forms at the cusp, which occurs at approximately $\lambda \simeq 4.77$ in the case $X_0 = 2.5$, $\gamma = 0.2$, and $\epsilon = 1/6$. \replace{}{The eventual formation of a current singularity is insensitive to the choice of these parameters.}}
    \label{fig:curr-sing}
\end{figure}

We now explore the space of Palumbo-like profiles using the quadratic ansatz for $\partial_x \Psi$ described above. We calculate the plasma profiles in terms of the solution of \eqref{eq:BTS_coupled_ODEs}, using equations \eqref{eq:u1_u2_def} and \eqref{eq:curr} to find the pressure and toroidal current (figures \ref{pressure} and \ref{fig:curr-sing}, respectively), and integrating 
\begin{equation}
    \iota^{-1} = -\frac{1}{2\pi}\oint \frac{dx}{v}
    \label{eq:iota}
\end{equation}
to find the rotational transform $\iota$. The integral over a flux surface in \eqref{eq:iota} is between two vertical asymptotes of the surface, i.e., two roots of the polynomial $v^2$. \replace{}{Such real roots are guaranteed to exist near the magnetic axis because we have enforced regularity of $\Psi$ in our choice of initial conditions; see Appendix \ref{app:Palumbo} for more details.}Thus the integral can be efficiently approximated using Chebyshev-Gauss quadrature \citep{abramowitz+stegun} or written explicitly in terms of elliptic integrals \citep{gradshteyn2007}.  

In the limit of small $\lambda$ \replace{(equivalently, as we approach the magnetic axis)}{(using the near-axis asymptotic behavior of \eqref{eq:BTS_coupled_ODEs} described in Appendix \ref{app:Palumbo})}, $v$ becomes
\begin{equation}
    v(s) \simeq X_0^2 \sqrt{-s^2(s-1)^2 + \gamma s^2/X_0^2},
\end{equation}
with $s = -x/X$. The radicand has a double root at $s = 0$ and two single roots
\begin{equation}
    s_{\pm} = 1\pm\sqrt{\frac{\gamma}{X_0^2}}.
\end{equation}
At positive $\lambda$, the degeneracy is split, so the double root becomes a pair $\{\lambda s_1, \lambda s_2\}$, and the other two roots gain an $\mathcal{O} (\lambda)$ correction. \replace{}{Substituting the asymptotic relations from Appendix \ref{app:Palumbo}, $s_1$ and $s_2$ are real if $X_0^2 > \gamma$. We will see that this is equivalent to the requirement that the on-axis rotational transform be real.} \replace{But in}{In} the limit of small $\lambda$, as we integrate between the roots near zero, these corrections can be neglected:
\begin{align}
    \iota^{-1}|_{\lambda \to 0} &= \lim_{\lambda \to 0}\frac{1}{2\pi X}\oint \frac{ds}{\sqrt{-(s-\lambda s_1)(s-\lambda s_2)(s-(1+\sqrt{\frac{\gamma}{X_0^2}}))(s-(1-\sqrt{\frac{\gamma}{X_0^2}}))}}\\
    &= \frac{1}{2\pi X_0}\frac{1}{\sqrt{1 - \frac{\gamma}{X_0^2}}}\lim_{\lambda \to 0}\oint \frac{ds}{\sqrt{-(s-\lambda s_1)(s - \lambda s_2)}}\\
    & = \frac{1}{X_0 \sqrt{1 - \frac{\gamma}{X_0^2}}}
\end{align}
where the integral can be evaluated for any nonzero $\lambda$ and gives $\pm 2\pi$. Thus the on-axis rotational transform is simply
\begin{equation}
    \iota = \sqrt{X_0^2 - \gamma}.
\end{equation}

It is possible to evaluate the integral \eqref{eq:iota} away from the magnetic axis in terms of elliptic functions, though this representation still depends on the roots of a quartic polynomial whose coefficients are the solutions to a nonlinear set of ODEs. Section 3.145 in \replace{Gradshteyn and Rizhik}{} \citep{gradshteyn2007} gives
\begin{equation}
    \iota^{-1} = \frac{2 u_2}{\pi X \sqrt(p q)}F\left(\pi, \frac{1}{2} \sqrt{\frac{(p+q)^2+(\alpha - \beta)^2}{p q}} \right)
    \label{eq:ellipticIota}
\end{equation}
where, using their notation, 
\begin{equation}
\begin{aligned}
    p^2 &= (m-\alpha)^2 + n^2 \\
    q^2 &= (m-\beta)^2 + n^2,
\end{aligned}
\end{equation}
$\alpha$ and $\beta$ are two real roots of $v^2$,  $m\pm i n$ are the complex conjugate pair of roots of $v^2$, and $F$ is the incomplete elliptic function of the first kind. A similar expression holds when all four roots of $v^2$ are real.

\begin{figure}
    \centering
    \includegraphics[width=0.6\columnwidth]{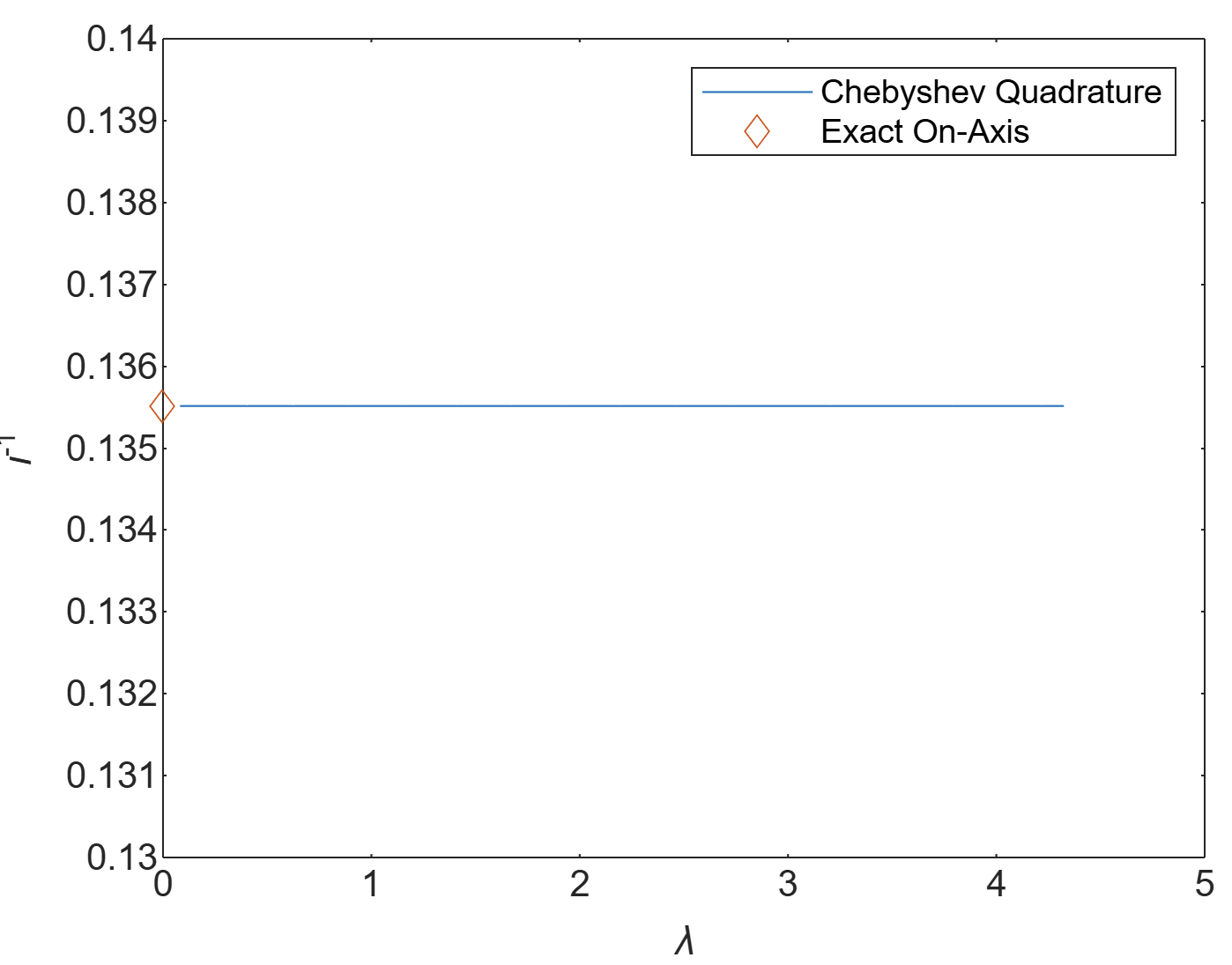}
    \caption{Integrating equation \eqref{eq:iota} with Gauss-Chebyshev quadrature shows a completely flat rotational transform throughout the plasma volume.}
    \label{fig:inv-iota}
\end{figure}

Further analysis of equation \eqref{eq:ellipticIota} is difficult\replace{}{, and we have been unable to prove that the magnetic shear vanishes, even with the given formula in terms of elliptic integrals, so we resort again to numerical methods}. Numerical integration using Gauss-Chebyshev quadrature, as in figure \ref{fig:inv-iota}, shows that the $\iota$ profile is completely flat. \textit{There is zero shear to order $\epsilon^2$ in these configurations}. \replace{}{Our model is, at least approximately, in the class of ``isochronous" dynamical systems, though not obviously of a form exhibited in the existing literature on isochronous systems \citep{calogero2008}.} We emphasize that this characteristic is robust to changes in parameters $\gamma$ and $X_0$ and completely absent from the models of Bishop and Taylor \citep{bishop_Taylor1986degenerate} or Hernandes and Clemente \citep{hernandes_Clemente2009ext_Palumbo}\replace{}{, both of which show the safety factor diverging at the separatrix}. However, it is consistent with the observed low shear seen in optimized QA and the analysis of weakly non-asixymmetric QA of Plunk \citep{plunk2020_near_axisymmetry_MHD}, which found that force balance makes the magnetic shear at most second order in the perturbation magnitude.

\subsection{Cusps and singularities}
We now consider the extreme limits of the equilibria produced by the system \eqref{eq:BTS_coupled_ODEs}. The singularities of this nonlinear set of ODEs correspond to a singular current profile and a corresponding flux surface cusp.

First, the limit of zero rotational transform requires $\gamma \to X_0^2$, which is a singular limit of the ODE initial conditions (see Appendix \ref{app:Palumbo}). It is apparent from figure \ref{D-shape} that this limit also creates extreme surface elongation near the axis, and D-shaped, positive-triangularity surfaces away from the axis.

At finite rotational transform, the ODEs \eqref{eq:BTS_coupled_ODEs} are well-behaved near the axis. The profiles eventually develop a singularity at some finite value of $\lambda$, just as two roots of $u$ and $v$ coincide. The shape of the flux surface is locally determined by the ODE
\begin{equation}
    \frac{d \xi}{dx} = \pm \frac{u}{v},
\end{equation}
so for simple roots of $u$ and $v^2$, $u\sim v^2 \sim (x-x_0)$ implies
\begin{equation}
    \xi \sim (x-x_0)^{3/2}.
    \label{eq:cusp}
\end{equation}
 The collision of roots of $u$ and $v^2$ also occurs at the magnetic axis, but there we can enforce the condition $\partial _x v^2 = 0$ to ensure local smoothness of $\Psi$. For a simple root of $v^2$, however, the flux surfaces lose analyticity, and a current singularity forms, as seen in figure \ref{fig:curr-sing}.

\begin{figure}
    \centering
    \includegraphics[width=0.65\columnwidth]{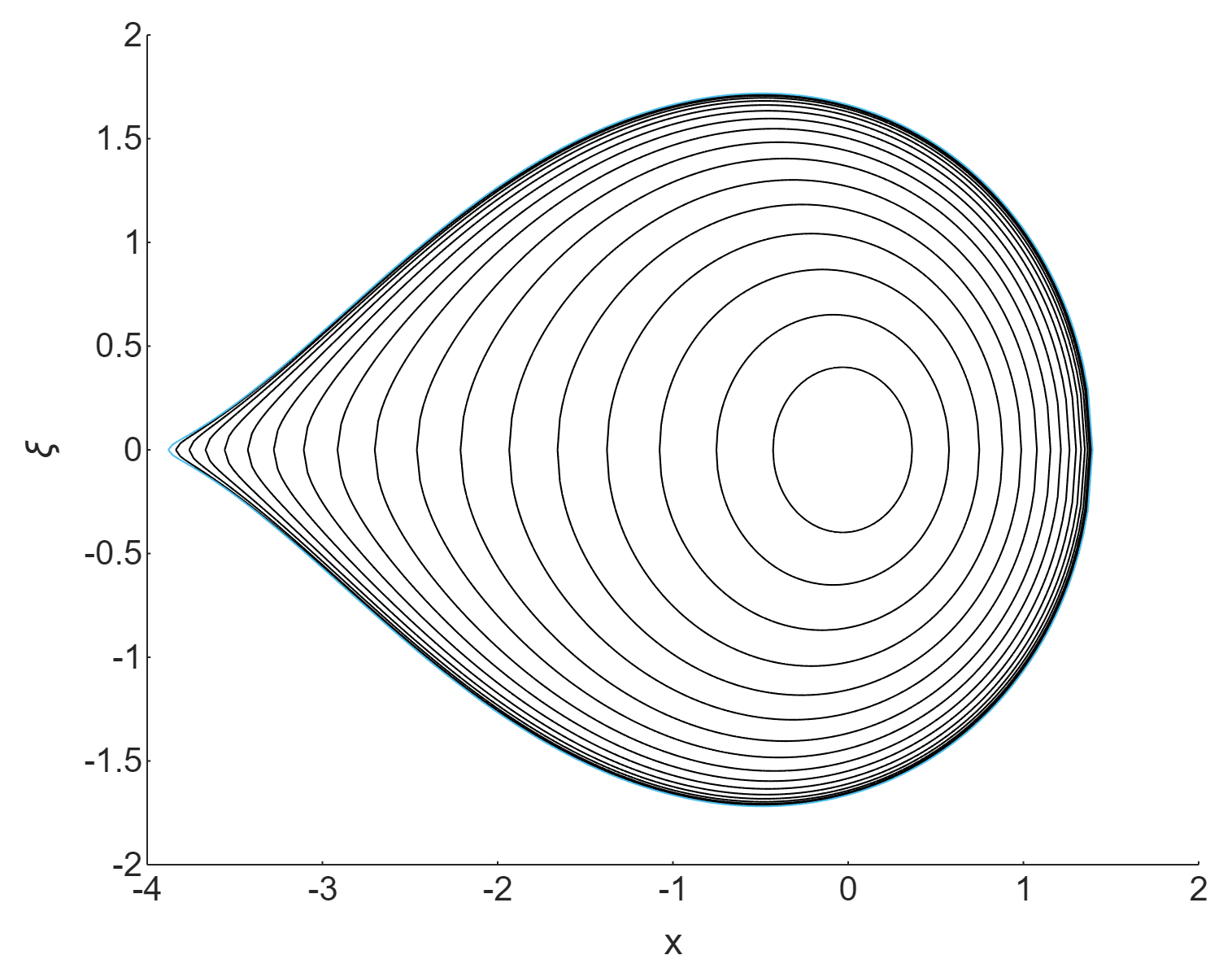}
    \caption{For most initial conditions, flux surfaces eventually form a cusp. Here $X_0 = 2.5$ and $\gamma = 0.2$.}
    \label{cusp-shape}
\end{figure}

Figure \ref{cusp-shape} shows the $\xi^2 \sim x^3$ cusp that forms instead of an X-point, in keeping with the proven non-existence of X-points for perfect QS with nonzero rotational transform\citep{rodriguez2021a}. Because the surface shape is independent of $\phi$, the cusp extends to an inboard ridge that moves in tandem with the magnetic axis. Such ridges are common features in optimized QS\citep{jorge2024single_stage}, but their behavior is poorly understood. \replace{}{Optimization efforts aimed at removing sharp features seem to merely push them out further from the magnetic axis, so they appear to be resilient features of QS \citep{paul2025}.}

The formation of an inboard poloidal field null at large pressure is consistent with experimental results at the poloidal beta limit, as observed in TFTR \citep{me_mauel_operation_1992}. Conversely, in the limit of zero rotational transform, X-points are allowed \replace{}{\citep{rodriguez2021a}}, which we do observe as flux surfaces push toward a D shape. Though it might appear to be an artifact of our particular choice of ansatz, the index $3/2$ in equation \eqref{eq:cusp} is much more general. Locally, this index is a hallmark of a cusp catastrophe or a saddle-node \replace{}{ or saddle-center} bifurcation \citep{arnold1984, meiss2007}\replace{}{, though we emphasize that no bifurcation \textit{per se} occurs in our model, as there is not sufficient freedom in parameter choice to permit the cusp to bifurcate into multiple critical points}. The $3/2$ cusp is one of only a handful of normal forms for catastrophes. \replace{}{See Appendix \ref{app:Palumbo} for a comparison of our model to another well-known Hamiltonian system that exhibits cusp formation for some choice of parameters.}

\begin{figure}
    \centering
    \includegraphics[width=0.65\columnwidth]{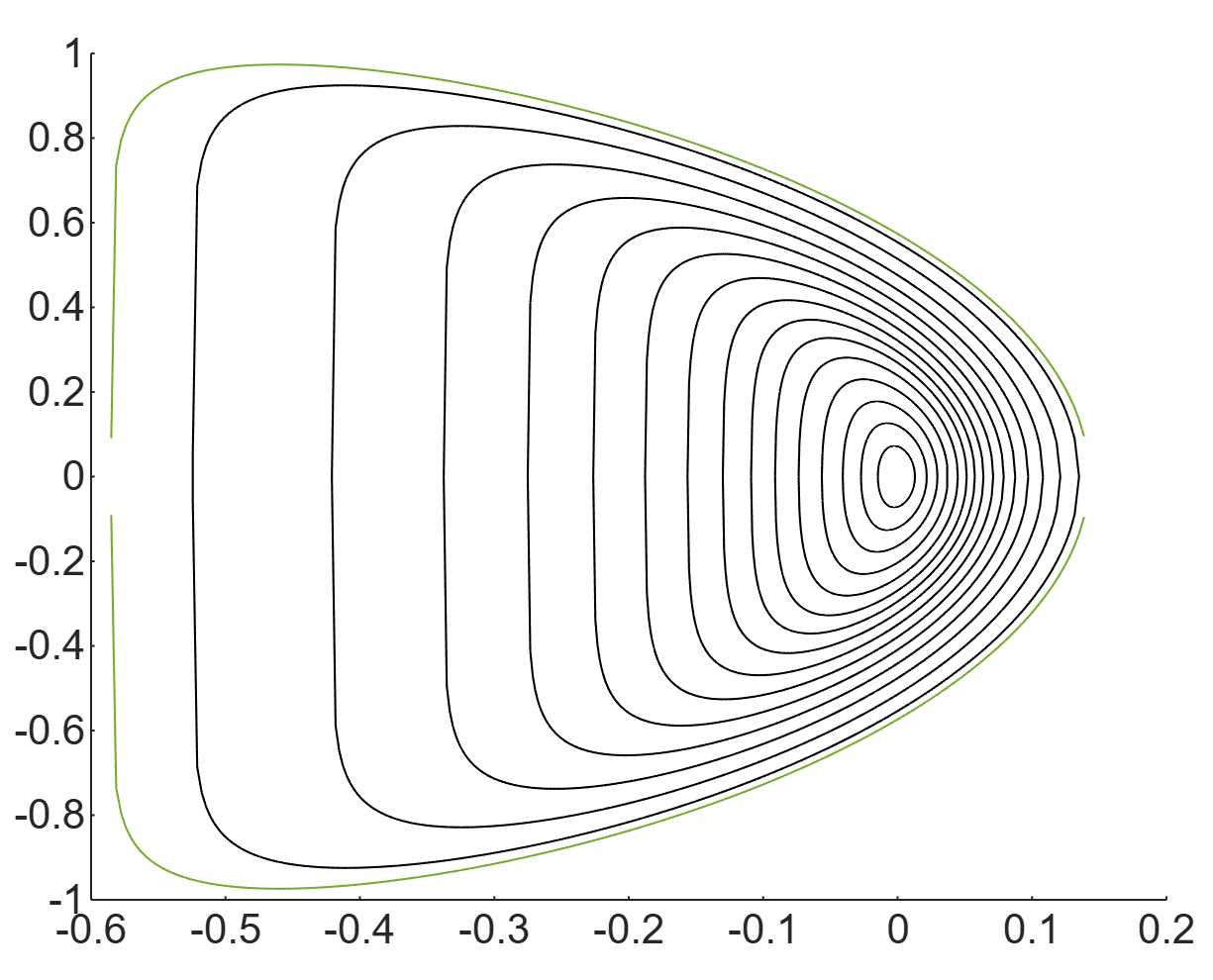}
    \caption{A D-shaped flux surface forms when the on-axis rotational transform is small enough. Here $X_0 = 1.5$ and $\gamma = 2.17$.}
    \label{D-shape}
\end{figure}

\begin{figure}
    \centering
    \includegraphics[width=0.6\columnwidth]{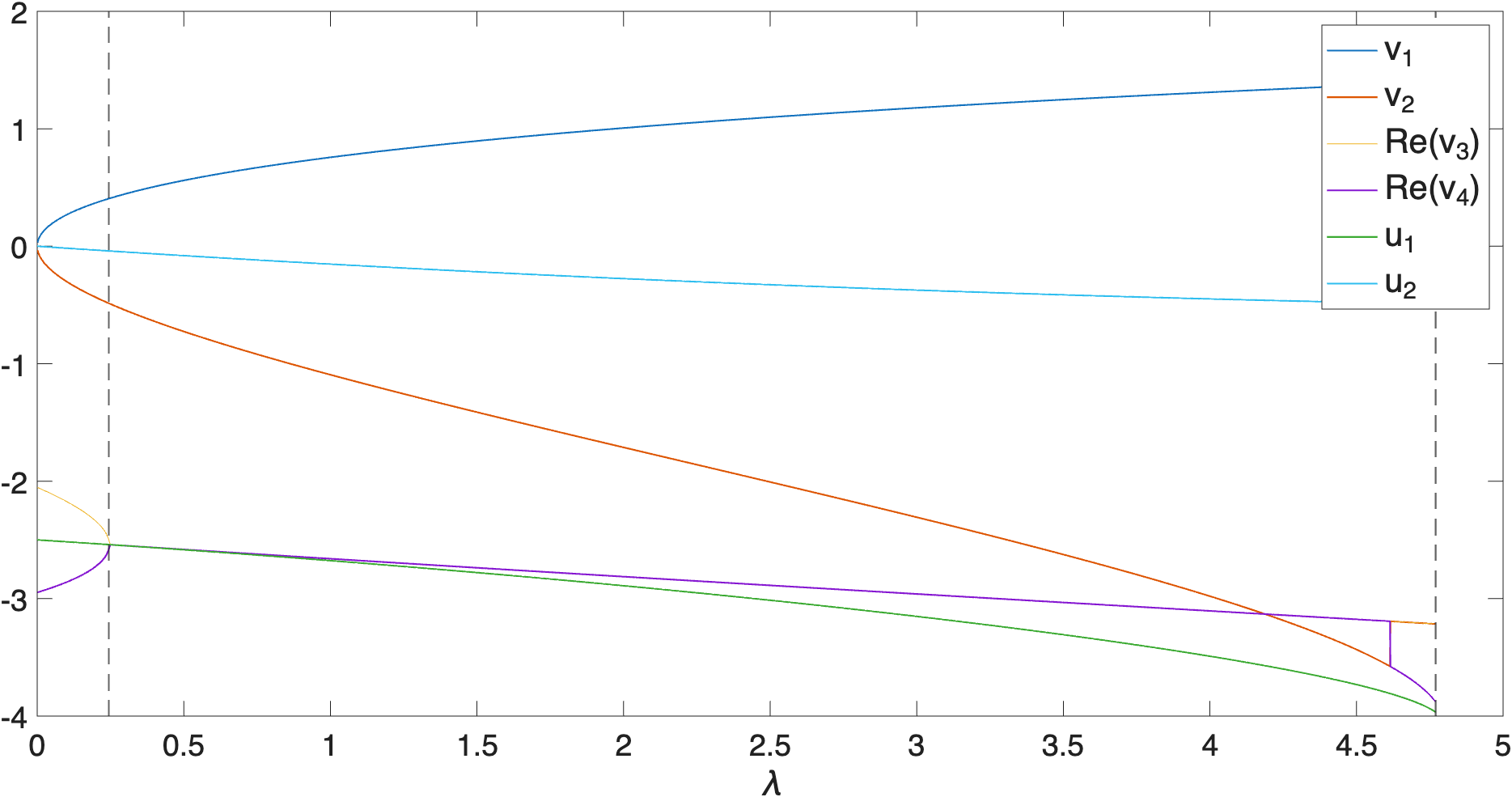}
    \caption{Roots of $u$ and $v$ as functions of the flux label. At the removable singularity $v_3=v_4=u_1$, but this $x$-value is outside the flux surface, which is delimited by $v_1$ and $v_2$. The cusp forms as $v_2$ approaches $u_1$. \replace{}{These two root collisions are indicated by vertical dashed lines.}}
    \label{fig:roots}
\end{figure}

In figure \ref{fig:roots}, we see that the flux label corresponding to the current singularity and cusp formation coincides with the collision of roots of $u$ and $v$, which occurs as the denominator of \eqref{eq:BTS_coupled_ODEs} vanishes. There is an earlier, removable singularity (here, at $\lambda \simeq 0.22$) as the numerator also vanishes. With sufficient resolution, the integration can be continued past the first singularity, whereas the latter singularity represents a breakdown of flux surface analyticity and an infinite (albeit integrable) current density.

As observed above, in the large-aspect-ratio limit, QA configurations must satisfy 
\begin{equation}
    \left(\frac{dB}{dl}\right)^2\sim \left(\frac{\partial \Psi}{\partial y}\right)^2 \sim \text{quartic or cubic in }x
\end{equation}
so $v \sim (x-x_0)^{1/2}$ near any simple root $x_0$ of $v$. Therefore the local form $y^2 \sim x^3$ is the generic behavior for a cusp in small-$\epsilon$, QA equilbria. \replace{}{See Appendix \ref{app:geom} for a more detailed discussion of the geometry of flux surfaces near the cusp.}

Because we are free to choose any nonzero value of $\iota$ throughout the plasma volume and thus in particular at the cusp, we emphasize that $\iota$ is not necessarily rational at the cusp, further distinguishing the cusp from a standard X-point. This family of solutions may perhaps serve as a prototypical class for the study of nonresonant divertors, which have recently been of considerable interest in the stellarator community\replace{}{ \citep{punjabi2020, garcia2023, davies2025}}.

\replace{Away from the cusp, a root of $v^2$ that does not coincide with a root of $u$ gives the principal curvatures $\kappa_1\sim -\epsilon (x-x_0)^{-1/2}$ and $\kappa_2\sim (x-x_0)^{1/2}$, so the Gaussian curvature $\mathcal{K}=\kappa_1\kappa_2$ is negative and $O(\epsilon)$, but $\mathcal{K}$ changes sign as the cusp forms since $\kappa_1\sim \epsilon (x-x_0)^{-1/2}$ when $u\sim (x-x_0)$. By continuity, there is thus some value of $\Psi$ at which $\mathcal{K}\to0$ near the cusp, right as the QS-HBS model breaks down. At the same time, the dominant toroidal field decays as $1/R$ and the field line is forced to lie on a ridge that has the same $O(\epsilon)$ deviation from circularity as the magnetic axis, so the magnetic field strength along the ridge only varies at $O(\epsilon^2)$. Therefore the field becomes locally isodynamic at the ridge. The connection between ridge formation in the last closed flux surface and zeros of the Gaussian curvature is the subject of ongoing research, to be presented in an upcoming publication. That QS should be unattainable when the Gaussian curvature vanishes is reminiscent of the fact, familiar from near-axis theory, that the axis curvature may not vanish in QA. See appendix \ref{app:geom} for details of calculations of geometric quantities.}{}

\section{Three-Dimensional Equilibria}
\label{sec:axis_pert}
\begin{figure}
    \centering
    \includegraphics[width=0.7\columnwidth]{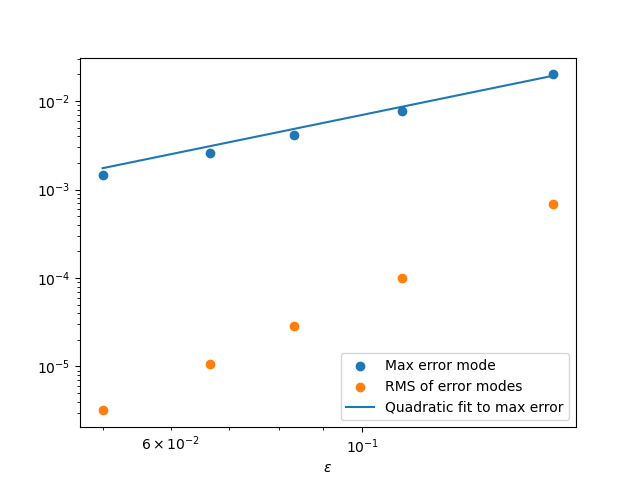}
    \caption{Maximum QS-breaking Boozer amplitude and RMS norm of QS-breaking modes, fitted against $\epsilon^2$ for comparison.}
    \label{fig:qs-vert-error}
\end{figure}

\begin{figure}
    \centering
    \includegraphics[width=0.85\columnwidth]{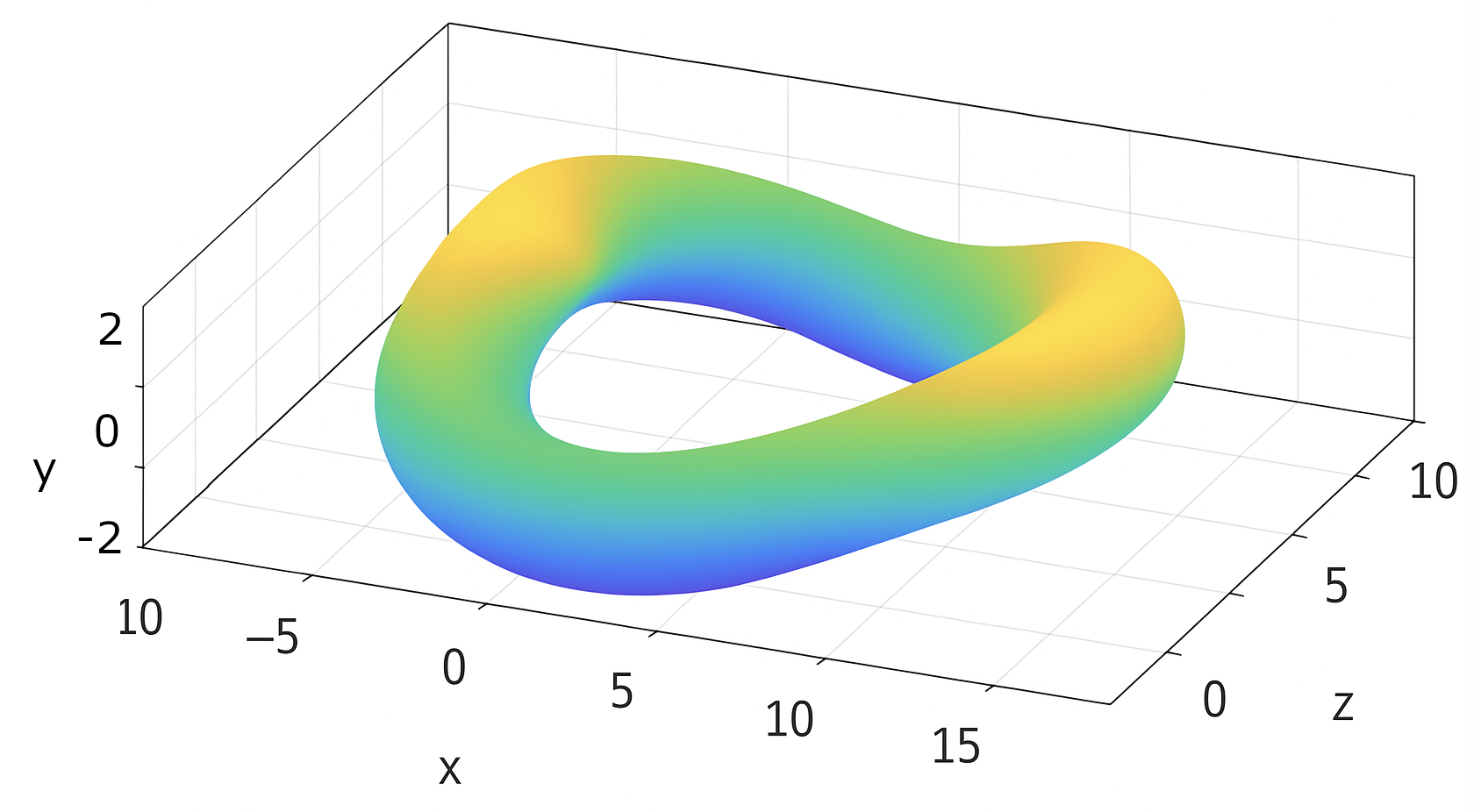}
    \caption{Plasma boundary with $m=2$ purely vertical axis perturbation. Color indicates $y$-value for visual clarity. \replace{}{Here global Cartesian coordinates are used, unrelated to the local coordinates in the rest of this work.}}
    \label{fig:vmec}
\end{figure}

\replace{}{In this section we explore permissible axis shaping and the associated tradeoffs between torsional rotational transform and QS error. The generalized QS-HBS model decouples flux surface shape, which is independent of the toroidal angle, from the magnetic axis. Using the SIMSOPT package BOOZ\_XFORM \citep{landreman2021-simsopt} to calculate the QS error, we see that a purely vertical, $n=2$ perturbation gives excellent quasisymmetry proportional to $\epsilon^2$, as shown in figure \ref{fig:qs-vert-error} and required by the asymptotic expansion underlying the QS-HBS model. Figure \ref{fig:vmec} illustrates one such vertical axis perturbation.}

Using the generalized HBS model recently described in \cite{Nikulsin_Sengupta_Jorge_Bhattacharjee_2024}, it can be shown that the axis perturbations may include some radial component, so long as the perturbation amplitude is $O(\epsilon)$ and the axis torsion is everywhere $O(\epsilon^2)$. Here we show that these conditions may be satisfied, generalizing a similar calculation of the axis torsion in \cite{Sengupta_Nikulsin_Gaur_Bhattacharjee_2024}. The torsional contribution to the total rotational transform \replace{}{\citep{mercier1974lectures}} then arises at $O(\epsilon)$, the same order as the current-driven part.

Let the axis be defined by
\begin{align}
    \bm{r}=R_0 (\bm{e_R} (1 + \mfg(\phi))+\mff(\phi)\bm{e_Z}).
\end{align}

Then the curvature is
\begin{equation}
    \kappa = \left\lvert \frac{d \bm{r}}{d\ell} \times \frac{d^2 \bm{r}}{d\ell^2} \right\rvert,
\end{equation}
and the torsion is 
\begin{equation}
    \tau = \frac{1}{\kappa^2} \frac{d \bm{r}}{d\ell} \times \frac{d^2 \bm{r}}{d\ell^2} \cdot \frac{d^3 \bm{r}}{d\ell^3}.
\end{equation}

It is convenient to define the arclength 
\begin{align}
    R_0 s(\phi) = \frac{d\ell}{d\phi} = \sqrt{\frac{d\bm{r}}{d\phi} \cdot \frac{d \bm{r}}{d \phi}} = R_0 \sqrt{(1 + \mfg)^2 + \mfg'^2 + \mff'^2},
\end{align}
so that the torsion becomes
\begin{equation}
    \tau = \frac{1}{\kappa^2}\frac{1}{R_0^6 s^6}\frac{\partial \bm{r}}{\partial \phi} \times \frac{\partial^2 \bm{r}}{\partial \phi^2} \cdot \frac{\partial^3 \bm{r}}{\partial \phi^3},
\end{equation}
in which we have excluded those parts of the derivatives of $\bm{r}$ that must vanish upon taking the triple product. 

Then the torsional part of the rotational transform is 
\begin{equation}
    \iota_{\tau} = \frac{1}{2\pi}\oint \mathrm{d}\phi s(\phi)\frac{\mathcal{A}}{\mathcal{B}},
    \label{eq:iota-tau}
\end{equation}
where the integrand is split for legibility into
\begin{equation}
\begin{aligned}
    \mathcal{A} = \mff'''&\left((1+\mfg)^2 - \mfg''(1+\mfg) + 2 \mfg'^2\right) - (1+\mfg)\left(\mff' \mfg'' - \mfg' \mff'' - \mff' (1+\mfg)\right) \\
    +& \left(\mff'' (1+\mfg) - 2 \mff' \mfg'\right)(\mfg''' - 3 \mfg')
\end{aligned}
\end{equation}
and
\begin{equation}
\begin{aligned}
    \mathcal{B} = & \left((1+\mfg)^2 - \mfg'' (1+\mfg) + 2 \mfg'^2\right)^2 + \left(\mff'' (1+\mfg)-2 \mff' \mfg'\right)^2  \\
    &+  \left(\mff' \mfg'' - \mfg' \mff'' - \mff' (1+\mfg) \right)^2
\end{aligned}
\end{equation}

In the work that introduced the HBS model, perturbations were restricted to purely vertical displacements, i.e., $\mfg \equiv 0$, and the integrated axis torsion was $O(\epsilon^3)$ \citep{Sengupta_Nikulsin_Gaur_Bhattacharjee_2024}. Our more complicated expression \eqref{eq:iota-tau} does reduce to the torsion found previously, \textit{viz.}, 
\begin{equation}
    \tau_{\mfg \to 0} = \frac{1}{R_0}\frac{\mff''' + \mff'}{1 + \mff''^2 + \mff'^2}.
\end{equation}
Expanding in powers of $\epsilon$ and setting $\mff = \epsilon \mff_1$, this expression gives
\begin{equation}
    \iota_{\tau, \mfg \to 0} = -\frac{1}{4 \pi}\oint \mathrm{d}\phi \mff_1'^3,
\end{equation}
different from the $\mff_1' \mff_1''^2$ integrand reported in previous work. 

\begin{figure}
    \centering
    \includegraphics[width=0.6\columnwidth]{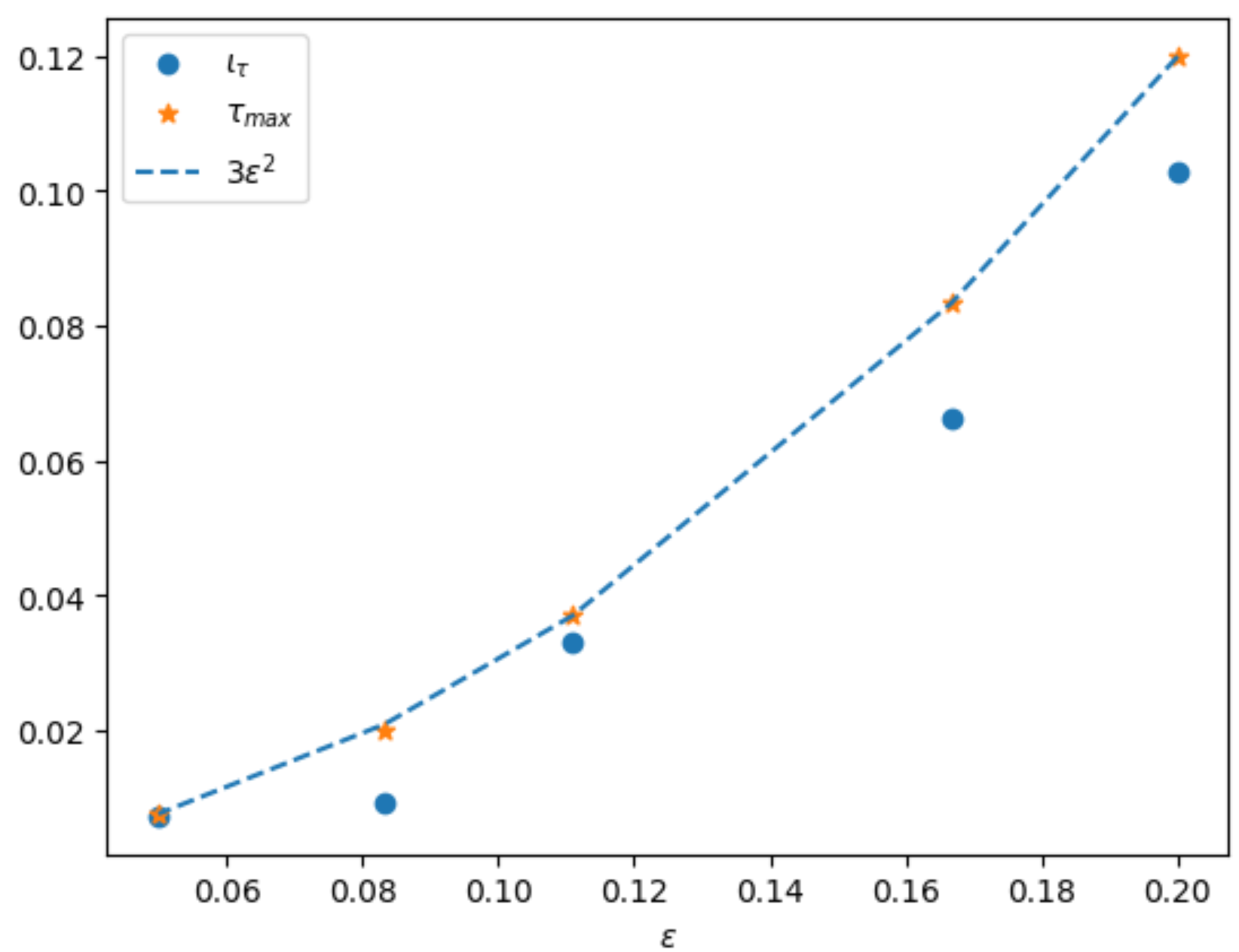}
    \caption{Numerically optimized rotational transform for two-mode perturbations ($n=1$ and $m=2$) across a range of aspect ratios, subject to the constraint $\tau_{max}\leq3\epsilon^2$. The maximum attainable $\iota_{\tau}$ is strongly correlated with the maximum allowable torsion. }
    \label{fig:optim-iota}
\end{figure}

Allowing for nonzero $\mfg$ makes $\iota_{\tau}$ formally $O(\epsilon^2)$, assuming $\mff$ and $\mfg$ are both $O(\epsilon)$ quantities. For judiciously chosen $\mff$ and $\mfg$, $\iota_{\tau}$ can be a significant contribution to the total rotational transform. For instance, taking $\mff$ and $\mfg$ to be sums of two harmonics, i.e., 

\begin{equation}
\begin{aligned}
    \mff(\phi)& = f_{c, m} \cos(m \phi) + f_{c,n} \cos(n \phi)+f_{s, m} \sin(m \phi) + f_{s,n} \sin(n \phi)\\
    \mfg(\phi) &= g_{s,m} \sin(m \phi) + g_{s,n} \sin(n \phi)+g_{c,m} \cos(m \phi) + g_{c,n} \cos(n \phi)\\
\end{aligned}
\end{equation}
 and choosing a range of aspect ratios $\epsilon^{-1} \in [5, 20]$, we numerically maximize $\iota_{\tau}$ subject to the (somewhat arbitrary) constraints $|a_{n,m}, b_{n,m}|\leq 3\epsilon$ and $\max\{\tau(\phi)\}\leq 3 \epsilon^2$. Keeping only $n=1$ and $m=2$, figure \ref{fig:optim-iota} shows that the axis contribution to the rotational transform can be significant, with the caveat that the axis torsion, although formally $O(\epsilon^2)$, will increase the quasisymmetry error.

\section{Conclusion}
We have constructed a novel class of solutions to the QS-GSE that arises in the limit of large-aspect-ratio, nearly-axisymmetric quasiaxisymmetry. These solutions exhibit surprising qualitative similarities to QA equilibria at finite aspect ratio, including negligible shear and the formation of surface cusps as the plasma volume approaches its maximum. \replace{}{The free parameters $X_0$ and $\gamma$ in this model primarily allow one to modulate the on-axis rotational transform, which is always constant throughout the plasma volume, and the total volume of the plasma, which is determined by singularity formation in the governing ODEs.}

We also note that the condition of weak non-axisymmetry means the Palumbo-QS-HBS model is a perturbed tokamak, so many features found in QS-HBS have well-known analogues in the tokamak literature. The QS-HBS ordering has $\beta = O(\epsilon)$ and $B_{\theta}/B_{\phi} = O(\epsilon)$, so 
$\beta_{pol} = O(1/\epsilon)$ and $\epsilon \beta_{pol}\sim 1$, a parameter regime that was explored in TFTR, which found that a poloidal field null forms on the inboard side at the poloidal beta limit \citep{me_mauel_operation_1992}. Furthermore, the slender D-shaped flux surfaces in QS-HBS at $\beta \sim \epsilon $ are reminiscent of Cowley's high-$\beta$ equilibria, which exhibit D-shaped surfaces and extreme Shafranov shift at $\epsilon \beta_{pol}\gg1$ \citep{cowley1991}. However, vanishing shear and the local breakdown of analyticity of $\Psi$ as a cusp (rather than an X-point) forms are not found in the aforementioned tokamak equilibria, providing suggestive, albeit inconclusive, evidence that these features are inherent to quasiaxisymmetry. In addition to these physical insights into QA, we hope that our Palumbo-like equilibria and other semi-analytical solutions to the QS-GSE will prove to be practically useful as hot starts in QA optimization.

\begin{acknowledgments}
The authors would like to thank S. Buller, E.J. Paul, G. Plunk, and M. Zarnstorff for their generous insights.

This work is supported by the DoE grant HiFiStell (until March 31, 2025) and the Simons Foundation Collaboration on Hidden Symmetries and Fusion Energy. 

\end{acknowledgments}

%\section{Appendices}
\appendix

\section{Higher-Degree Flux Ans{\"a}tze}
\label{app:overdet}
Here we show that the method of section \ref{sec:Extended_Palumbo} can be applied to a cubic polynomial ansatz for $\eta$, but for any polynomial of degree $4$ or greater the resulting system of ODEs is overdetermined.

First, assuming a cubic ansatz
\begin{equation}
    \eta(x, \Psi) = \eta_3 x^3 + \eta_2(\Psi) x^2 + \eta_1(\Psi)x + \eta_0(\Psi),
\end{equation}
where again the coefficient of the highest-degree term is taken to be constant so that $u$ is also cubic, we find the following set of ODEs:

\begin{subequations}
\begin{align}
2 u_0 u_1 - \eta_1 - u_0 \eta_0' + 2 u_0' \eta_0 &= 0 \\
2 u_1^2 + 4 u_0 u_2 - 2\eta_2 - u_0 \eta_1' - u_1 \eta_0' + 2 u_0' \eta_1 + 2 u_1' \eta_0 &= 0 \\
6 u_1 u_2 - \frac{2}{3} \eta_2' u_0 - 3 \eta_3 - \eta_2' u_0 + 2\eta_2 u_0' - \eta_1' u_1 + 2\eta_1 u_1' - \eta_0' u_2 + 2 \eta_0 u_2' &= 0 \\
4 u_2^2 - \eta_2' u_1 + \frac{1}{6} \eta_0' \eta_2' + 2 \eta_1 u_2' + 2 \eta_3 u_0' - \eta_2' u_1 + 2\eta_2 u_1' - \eta_1' u_2 &= 0 \\
\frac{1}{6} \eta_1' \eta_2' - \frac{4}{3} \eta_2' u_2 + 2 \eta_3 u_1' - \eta_2' u_2 + 2\eta_2 u_2' &= 0 \\
\frac{5}{18}\left(\eta_2'\right)^2 + 2 \eta_3 u_2' &= 0
\end{align}
\end{subequations}
As $\eta_3$ and $u_3$ are constants, we have six equations in six unknowns. However, if $\eta$ is quartic, and we take $\eta_4$ to be a nonzero constant, the equality
\begin{equation}
    u = u_0 - x H' - x^2 \beta' - \frac{1}{2}\int dx \left[\eta_0' + x \eta_1' + x^2 \eta_2 ' + x^3 \eta_3'\right]
\end{equation}
gives trivial relations between $u_4$ and $\eta_3$. Once again applying the consistency condition \eqref{eq:consistency_eta_u}, we match terms from $O(x^0)$ up to $O(x^8)$, yielding nine equations for the seven unknowns $u_0$ up to $u_2$ and $\eta_0$ up to $\eta_3$. By a similar argument, polynomial ansatzes of degree four or greater always generate overdetermined systems. 

\section{Details of the extended Palumbo solution}
\label{app:Palumbo}

%\subsection{Closure of the flux surfaces: Near-axis behavior of the ODE system \eqref{eq:BTS_coupled_ODEs}}
The ODE system \eqref{eq:BTS_coupled_ODEs} is highly nonlinear and can have solutions that are not regular or divergent near the axis. Thus, a proper regularity boundary condition must be imposed near the axis. Moreover, as discussed in Appendix A of \cite{bishop_Taylor1986degenerate}, the closure condition for the flux surfaces holds in the entire volume if it holds near the axis. Therefore, we must carefully look at the near-axis behavior of the system \eqref{eq:BTS_coupled_ODEs}. 

The expression for $\xi$ given in \eqref{eq:xi_exp} suggests the following new variables
\begin{align}
\overline{P}=\frac{P}{X^4}, \quad \overline{Q}=\frac{Q}{X^3}, \quad \overline{\Gamma}=\frac{\Gamma}{X^2}, \quad 
\overline{\gamma}=\frac{\gamma}{X^2}, \quad 
\overline{\xi}=\frac{\xi}{X}, \nonumber\\
    \text{such that} \quad \quad  \overline{\xi} = \int ds \frac{\overline{\Gamma}+s(s-1)}{\sqrt{\lbr \overline{P}-s \overline{Q}+s^2 \overline{\gamma} \rbr-\lbr s(s-1)+\overline{\Gamma}\rbr^2 }}.  \label{eq:new_bar_variables}
\end{align}
%For simplicity, we shall now proceed with only the case $\gamma=0$, corresponding to the constant $\eta_2=0$.

Near the magnetic axis $(x=0,\Psi=0)$, $\eta=|\dlr \Psi|$ must vanish. Since $\eta=u^2+v^2$, we must have both $u=v=0$ at $x=0,\Psi=0$. Furthermore, flux surfaces of elliptic or circular cross-sections can exist near the axis provided that $v^2$ is approximately a quadratic with two real roots, the roots coinciding on the magnetic axis. Therefore, $\del_x v^2$ must vanish on the axis. Thus, we get three conditions 
\begin{align}
u=0, \quad v=0, \quad \del_x v^2 =0 \quad \text{at} \quad (x=0, \Psi=0).
    \label{eq:NAE_Clemente_conditions}
\end{align}
From \eqref{eq:eta_ansatz}, \eqref{eq:u_form2}, and \eqref{eq:PQXetc_def} it follows that \eqref{eq:NAE_Clemente_conditions} implies
\begin{align}
    P(0)=0, \quad \Gamma(0) =0 ,\quad Q(0)=0
    \label{eq:PGamaQ_0}
\end{align}

The solution to \eqref{eq:BTS_coupled_ODEs} is very well approximated by its near-axis approximations of the form
\begin{align}
X &=X_0+\lambda X_1 + \lambda^2 X_2+O(\lambda^3), \nonumber\\
P &=  \lambda P_1 + \lambda^2 P_2+ O(\lambda^3), \quad \Pb= \lambda \Pb_1 + \lambda^2 \Pb_2+ O(\lambda^3) \nonumber\\
\Gamma &= \lambda \Gamma_1 + \lambda^2 \Gamma_2+ O(\lambda^3),\quad \Gammab= \lambda \Gammab_1 + \lambda^2 \Gammab_2+ O(\lambda^3),\\
Q &= \lambda Q_1 + \lambda^2 Q_2+ O(\lambda^3), \quad \Qb= \lambda \Qb_1 + \lambda^2 \Qb_2+ O(\lambda^3) \nonumber
\end{align}
where $X_i, P_i...$ etc are coefficients given by:
\begin{align}
X_1&=\frac{1}{2}\lbr \frac{1}{X_0^2}+\frac{3}{X_0^2-\gamma}\rbr, \quad  X_2=-\frac{3 \gamma (8 X_0^4 - 7 X_0^2 \gamma + 
   2 \gamma^2)}{16 X_0^5 (X_0^2 - \gamma)^3} \nonumber\\
   P_1&=2 X_0-\frac{2 \gamma }{X_0}, \quad P_2= \frac{\left(4 X_0^2-3 \gamma \right) \left(\gamma ^2+10 X_0^4-2 \gamma  X_0^2\right)}{4 X_0^4 \left(X_0^2-\gamma \right)^2} \nonumber\\
    \Gamma_1&=\frac{  \left(\gamma +2 X_0^2\right)}{2X_0( X_0^2-\gamma)},\quad \Gamma_2=\frac{2 \gamma ^3-8 X_0^6-3 \gamma ^2 X_0^2}{16 X_0^4 \left(X_0^2-\gamma \right)^3}  \label{eq:coeff_Pi_Xi_etc}\\
    Q_1&=  \left(\frac{\gamma }{X_0^2}+\frac{3 \gamma }{X_0^2-\gamma }+4\right),\quad Q_2=\frac{3 \left(\gamma ^2 \left(2 \gamma ^2+8 X_0^4-7 \gamma  X_0^2\right)\right)}{8 \left(X_0^5 \left(X_0^2-\gamma \right)^3\right)}\nonumber
\end{align}
The convergence of the series in $\lambda$ clearly requires $X_0>1, \gamma\ll X_0^2$. The limit $\gamma \to X_0^2$, in which the rotational transform vanishes, is singular. The quantities $\Pb,\Qb,\gammab,\Gammab$ can be similarly expanded as
\begin{align}
\Pb &=\frac{2 \lambda  \left(X_0^2-\gamma \right)}{X_0^5}+\frac{\lambda ^2 \left(13 \gamma ^3-24 X_0^6+106 \gamma  X_0^4-86 \gamma ^2 X_0^2\right)}{4 X_0^8 \left(X_0^2-\gamma \right)^2}+O\left(\lambda ^3\right) \nonumber\\
\Qb &=\frac{\lambda  \left(\frac{3 \gamma }{X_0^2-\gamma }+\frac{\gamma }{X_0^2}+4\right)}{X_0^3}-\frac{3 \lambda ^2 \left(-2 \gamma ^4+64 X_0^8-80 \gamma  X_0^6+8 \gamma ^2 X_0^4+13 \gamma ^3 X_0^2\right)}{8 \left(X_0^8 \left(X_0^2-\gamma \right)^3\right)}+O\left(\lambda ^3\right)\nonumber\\
\Gammab &=\frac{\lambda  \left(\gamma +2 X_0^2\right)}{X_0^2 \left(2 X_0^3-2 \gamma  X_0\right)}-\frac{3 \lambda ^2 \left(2 \gamma ^3+24 X_0^6-16 \gamma  X_0^4-7 \gamma ^2 X_0^2\right)}{16 \left(X_0^6 \left(X_0^2-\gamma \right)^3\right)}+O\left(\lambda ^3\right)\\
\gammab &= \frac{\gamma }{X_0^2}-\frac{\gamma  \lambda  \left(\frac{3}{X_0^2-\gamma }+\frac{1}{X_0^2}\right)}{X_0^3}+\frac{3 \gamma  \lambda ^2 \left(11 \gamma ^2+32 X_0^4-40 \gamma  X_0^2\right)}{8 X_0^6 \left(X_0^2-\gamma \right)^3}+O\left(\lambda ^3\right).\nonumber
    \label{eq:Pbar_Qbar_etc_NAE}
\end{align}

\section{Geometry of the Extended Palumbo Solution}
\label{app:geom}
In this section we calculate in detail some geometric properties of axisymmetric extended Palumbo equilibria. For clarity we explicitly include a factor of $\epsilon$ in the poloidal field component, rather than subsuming this factor into the definition of $u_2$.

\replace{}{The magnetic field is, to first order in $\epsilon$,}
\replace{}{\begin{equation}
    \mathbf{B} = B_0 \nabla \phi + \epsilon \hat{\phi}\times\nabla_\perp \Psi,
\end{equation}
}
\replace{T}{so t}he unit vector tangent to magnetic field lines is, in cylindrical coordinates, 
\begin{equation}
    \hat{\boldsymbol{b}} = \frac{B_0}{B(1 + \epsilon x)}\hat{\mathbf{\phi}}+\frac{\epsilon}{B} \left(v \hat{\boldsymbol{R}}- u \hat{\boldsymbol{Z}}\right)
\end{equation}
and the surface normal is
\begin{equation}
    \hat{\boldsymbol{n}} = \frac{u \hat{\boldsymbol{R}}+v \hat{\boldsymbol{Z}}}{\sqrt{u^2 + v^2}}.
\end{equation}

The geodesic curvature may be written as the triple product
\begin{equation}
    \kappa_g = \frac{d \hat{\boldsymbol{b}}}{dl}\cdot \left(\hat{\boldsymbol{n}}\times\hat{\boldsymbol{b}}\right).
    \label{eq:kappa_g}
\end{equation}

The field line curvature 
\begin{equation}
    \frac{d\hat{\boldsymbol{b}}}{dl} = -\epsilon \hat{\boldsymbol{R}} + O(\epsilon^2)
\end{equation}
so, inserting into equation \eqref{eq:kappa_g}, 
\begin{equation}
    \kappa_g = -\epsilon \frac{v}{|\nabla\Psi|}
\end{equation}
A flux surface cusp forms when $u$ and $v^2$ share a common, simple root at $x_0$. Inserting local behavior $u\sim x-x_0$ and $v\sim (x-x_0)^{1/2}$, we find
\begin{equation}
\begin{aligned}
    \kappa_g &\sim \frac{(x-x_0)^{1/2}}{\sqrt{(x-x_0)^2 + (x-x_0)}} \sim \frac{(x-x_0)^{1/2}}{(x-x_0)^{1/2}},
\end{aligned}
\end{equation}
so the geodesic curvature stays finite even as the flux surfaces form a cusp. The normal curvature is
\begin{equation}
    \kappa_n = \frac{\boldsymbol{\kappa}\cdot\nabla\Psi}{|\nabla \Psi|} = -\epsilon \frac{u}{|\nabla \Psi|} + O(\epsilon^2).
\end{equation}
which varies as $\kappa_n \sim (x-x_0)^{1/2}$ near the cusp, vanishing while the geodesic curvature stays finite.

To find the principal curvatures, we calculate the first and second fundamental forms of the outermost flux surface, parametrized in Cartesian coordinates as 
\begin{equation}
    \boldsymbol{r} = (R \cos\phi, R \sin \phi, Z)
\end{equation}
where $Z = Z (R, \Psi)$ is independent of $\phi$ in the axisymmetric case. Then the normal vector
\begin{equation}
\begin{aligned}
    \hat{\boldsymbol{n}}&= \frac{\boldsymbol{r}_{R}\times \boldsymbol{r}_{\phi}}{|\boldsymbol{r}_{R}\times \boldsymbol{r}_{\phi}|}\\
    &= \frac{1}{\sqrt{1 + \left(\partial_R Z\right)^2}}\left(-\partial_R Z\cos\phi , -\partial_R Z\sin\phi,1\right),
\end{aligned}
\end{equation}
the first fundamental form is
\begin{equation}
    \mathrm{I}_p = E dR^2 + G d\phi^2
\end{equation}
where
\begin{equation}
    \begin{aligned}
    E &= \partial_R \boldsymbol{r}\cdot\partial_R \boldsymbol{r}= \left(1 + \left(\partial_RZ\right)^2\right)\\
    G &= \partial_{\phi} \boldsymbol{r}\cdot\partial_{\phi} \boldsymbol{r}= R^2
    \end{aligned}
\end{equation}
and the second fundamental form is 
\begin{equation}
    \mathrm{I\!I}_p = L dR^2 + Nd\phi^2
\end{equation}
where
\begin{equation}
\begin{aligned}
    L&=\boldsymbol{r}_{RR}\cdot\hat{\boldsymbol{n}}
    =\frac{\partial_R^2Z}{\sqrt{1 + \left(\partial_RZ\right)^2}}\\
    N &= \boldsymbol{r}_{\phi\phi}\cdot\hat{\boldsymbol{n}}
    = \frac{R \partial_R Z}{\sqrt{1 + \left(\partial_RZ\right)^2}}.
\end{aligned}
\end{equation}
Since the first and second fundamental forms are both diagonal, we can read off the principal curvatures
\begin{equation}
\begin{aligned}
    \kappa_1 &= \frac{L}{\sqrt{EG}}
    =\frac{\partial_R^2 Z}{R\left(1 + \left(\partial_RZ\right)^2\right)}\\
    &\simeq\epsilon \frac{v\frac{du}{dx} - u\frac{dv}{dx}}{u^2 + v^2}
\end{aligned}
\end{equation}
and
\begin{equation}
    \begin{aligned}
    \kappa_2 &= \frac{N}{\sqrt{EG}}
    =\frac{R \partial_R Z}{R\left(1 +\left(\partial_RZ\right)^2\right)}\\
    &=\frac{uv}{u^2 + v^2}
\end{aligned}
\end{equation}
using $\partial_R Z\to \frac{d\xi}{dx}=\pm\frac{u}{v}$.

\replace{}{Away from the cusp, a root of $v^2$ that does not coincide with a root of $u$ gives the principal curvatures $\kappa_1\sim -\epsilon (x-x_0)^{-1/2}$ and $\kappa_2\sim (x-x_0)^{1/2}$, so the Gaussian curvature $\mathcal{K}=\kappa_1\kappa_2$ is negative and $O(\epsilon)$, but $\mathcal{K}$ changes sign as the cusp forms since $\kappa_1\sim \epsilon (x-x_0)^{-1/2}$ when $u\sim (x-x_0)$. By continuity, there is thus some value of $\Psi$ at which $\mathcal{K}\to0$ near the cusp, right as the QS-HBS model breaks down. At the same time, the dominant toroidal field decays as $1/R$ and the field line is forced to lie on a ridge that has the same $O(\epsilon)$ deviation from circularity as the magnetic axis, so the magnetic field strength along the ridge only varies at $O(\epsilon^2)$. Therefore the field becomes locally isodynamic at the ridge. The connection between ridge formation in the last closed flux surface and zeros of the Gaussian curvature is the subject of ongoing research, to be presented in an upcoming publication. That QS should be unattainable when the Gaussian curvature vanishes is reminiscent of the fact, familiar from near-axis theory, that the axis curvature may not vanish in QA.}

\section{Analogy to the Heavy Symmetric Top}
Here we quote results from Goldstein's exposition \citep{goldstein2002} (Section 5.7) of the heavy symmetric top with one point fixed, and elucidate parallels between that system and the modified Palumbo equilibria constructed in this work. Following Goldstein's notation, the time elapsed by an orbit $\mathcal{U}(t)$ takes the form of an elliptic integral
\begin{equation}
    t = \int_{\mathcal{U}(0)}^{\mathcal{U}(t)}\frac{d\mathcal{U}}{\sqrt{(1-\mathcal{U}^2)(\alpha - \beta \mathcal{U})-(b - a\mathcal{U})^2}}
    \label{eq:top}
\end{equation}
and the integrand is the square root of a cubic polynomial
\begin{equation}
    f(\mathcal{U}) = \beta \mathcal{U}^3 - (\alpha + a^2) \mathcal{U}^2 + (2 a b - \beta)\mathcal{U} + (\alpha -b^2)
\end{equation}
in the variable $\mathcal{U} = \cos \theta$, where $\theta$ is the angle between the vertical and the symmetry axis of the top. The comparison to equation \eqref{eq:iota} is immediate; in our case, $v$ is the square root of a quartic in $x$ instead of a cubic in $\mathcal{U}$, but a formal similarity is apparent.

\begin{figure}  
    \begin{subfigure}[b]{\linewidth}
        \centering
        \includegraphics[width=0.8\columnwidth]{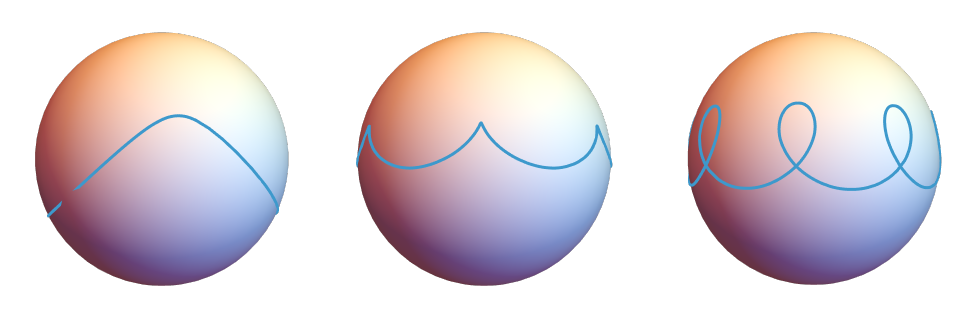}
    \end{subfigure}
    \begin{subfigure}[b]{\linewidth}
        \centering
        \includegraphics[width=0.85\columnwidth]{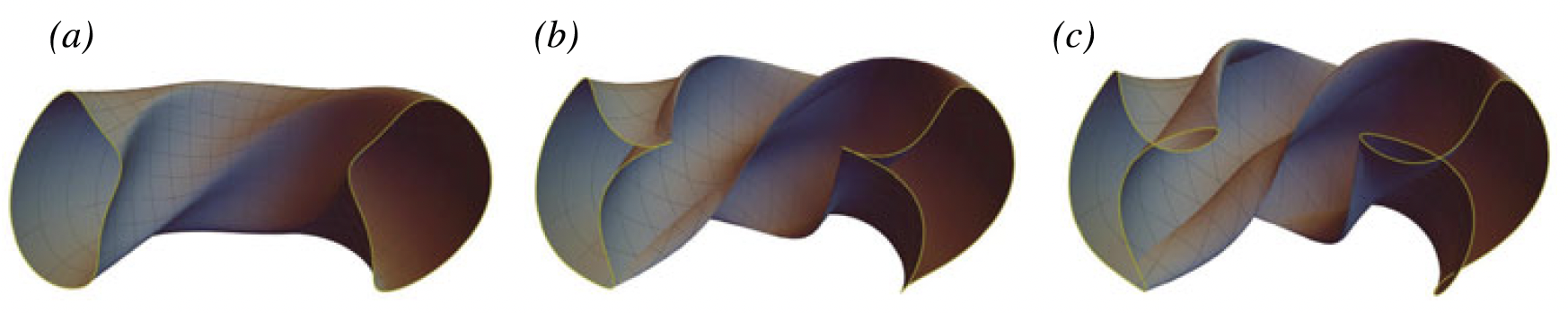}
    \end{subfigure}
    \caption{Top: Paths traced by a precessing, nutating heavy symmetric top in the cases of (a) smooth ($u'>u_2$), (b) cusped ($u'=u_2$), and (c) looped ($u_1< u'<u_2$) nutation. Bottom: Analogous magnetic smooth, cusped, and looped magnetic surfaces as the non-axisymmetric perturbation is increased. Reproduced from  \citet{plunk2020_near_axisymmetry_MHD} with permission.}
    \label{fig:top}
\end{figure}

The behavior of the top may be qualitatively understood from the root of the monomial
\begin{equation}
    g(\mathcal{U}) \equiv b - a\mathcal{U}
\end{equation}
in relation to the roots $\mathcal{U}_{1,2,3}$ of $f(\mathcal{U})$, which are all real. If $\mathcal{U}_1 < \mathcal{U}'< \mathcal{U}_2$, the tip of the top traces loops on the unit sphere; if $\mathcal{U}' > \mathcal{U}_2$, the top smoothly nutates without looping, as shown in figure \ref{fig:top}. Most interestingly, if $\mathcal{U}' = \mathcal{U}_1$ or $\mathcal{U}' = \mathcal{U}_2$, the loops contract and the locus develops cusps. In other words, the top locus develops cusps when $g(\mathcal{U})$, a polynomial of degree $1$, has a common root with $f(\mathcal{U})$, a polynomial of degree $3$. In the modified Palumbo case, $\partial_x \Psi$ is a polynomial of degree $2$ and $(\partial_{\xi}\Psi)^2$ is a polynomial of degree $4$, but just as for the top, a cusp forms when these two polynomials share a root. We also note that the sum 
\begin{equation}
    f(\mathcal{U})+g(\mathcal{U})^2 = (1-\mathcal{U}^2)(\alpha - \beta \mathcal{U})
\end{equation}
is analogous to $|\nabla_{\perp}\Psi|^2$, which is a quadratic in $x$ in the modified Palumbo case. In both models, when $g(\mathcal{U})$ or $u$ go to zero simultaneously with $f(\mathcal{U})$ or $v^2$, there is a qualitative change in the local behavior of the integrand in the $t$ (or $\iota^{-1}$) integrals.

A crucial difference is that the higher degree in the Palumbo case allows for a pair of complex roots, which may collide outside the domain of the flux surface, creating a removable singularity instead of a physically realized cusp.

%%%%%%%%%%%%%% Bibliography %%%%%%%%%%%%%%%%%%%%

\bibliographystyle{jpp}
\bibliography{plasmalit}

\end{document}